\documentclass[conference,a4paper,table]{IEEEtran}
\IEEEoverridecommandlockouts

\usepackage[T1]{fontenc} 
\usepackage{amsmath,amssymb,amsfonts}
\usepackage{algorithmic}
\usepackage{graphicx}
\graphicspath{{figures/}}
\usepackage{booktabs}
\usepackage{textcomp}
\usepackage{todonotes}
\usepackage{hyperref}

\def\BibTeX{{\rm B\kern-.05em{\sc i\kern-.025em b}\kern-.08em
    T\kern-.1667em\lower.7ex\hbox{E}\kern-.125emX}}

\usepackage{array}
\newcolumntype{H}{>{\setbox0=\hbox\bgroup}c<{\egroup}@{}}

\usepackage{xcolor}
\usepackage{tabularx}

\usepackage[backend=biber,
            style=numeric,
            natbib=true,
            doi=false,
            isbn=false,
            url=false,
            maxnames=1,
            maxcitenames=1,
            giveninits=true, 
            bibencoding=utf8]{biblatex}

\begin{document}

\title{Appeal prediction for AI up-scaled Images}

\author{%
\IEEEauthorblockN{Steve G\"oring, Rasmus Merten, Alexander Raake \\
\IEEEauthorblockA{Audiovisual Technology Group; Technische Universität Ilmenau, Germany\\
Email: [steve.goering, rasmus-leo-lukas.merten, alexander.raake]@tu-ilmenau.de}
}
}

\maketitle

\begin{abstract}
DNN- or AI-based up-scaling algorithms are gaining in popularity due to the improvements in machine learning.
Various up-scaling models using CNNs, GANs or mixed approaches have been published.
The majority of models are evaluated using PSRN and SSIM or only a few example images.
However, a performance evaluation with a wide range of real-world images and subjective evaluation is missing, which we tackle in the following paper.
For this reason, we describe our developed dataset, which uses 136 base images and five different up-scaling methods, namely Real-ESRGAN, BSRGAN, waifu2x, KXNet, and Lanczos.
Overall the dataset consists of 1496 annotated images.
The labeling of our dataset focused on image appeal and has been performed using crowd-sourcing employing our  open-source tool AVRate Voyager.
We evaluate the appeal of the different methods, and the results indicate that Real-ESRGAN and BSRGAN are the best.
Furthermore, we train a DNN to detect which up-scaling method has been used, the trained models have a good overall performance in our evaluation.
In addition to this, we evaluate state-of-the-art image appeal and quality models, here none of the models showed a high prediction performance, therefore we also trained two own approaches.
The first uses transfer learning and has the best performance, and the second model uses signal-based features and a random forest model with good overall performance.
We share the data and implementation to allow further research in the context of open science.
\end{abstract}


\section{Introduction} \label{sec::intro}
The recent developments in image processing focus on AI-based image enhancement and such methods are able to replace traditional signal-based approaches.
In general, AI-based methods are used for several image enhancement problems, for example, de-noising~\cite{laine2019high}, up-scaling~\cite{wang2021realesrgan,fu2022kxnet,dong2014learning,zhang2021designing}, image in-painting~\cite{pathak2016context}, or re-colorization~\cite{salmona2022deoldify}.
The used models to perform the tasks are typically based on deep neural networks (DNNs), e.g., GANs~\cite{wang2018esrgan}  (generative adversarial networks), U-net~\cite{ronneberger2015u}, or Autoencoders~\cite{chira2023image}.
The published models are usually evaluated using objective metrics, such as PSNR, or SSIM.
Instead of such objective metrics, which may have also their limitations, we focus in the following on subjective evaluation of up-scaling algorithms.
Because DNNs do not use traditional approaches for up-scaling, it can happen that the used method introduces completely new image parts or new types of distortions.
To properly evaluate such generated images and estimate which algorithm performs best or is most suitable for an application, human annotations are essential and required.
Therefore, it is needed to evaluate the appeal and quality of such generated images.
In general, it can be observed that there is a strong link between image appeal and image quality, especially when there are no explicit encoding distortions considered~\cite{goering2023aiquality}.
Furthermore, as shown in~\cite{goering2023ai} AI-generated content may be harder to analyze with current models and features for image appeal.
These are the reasons why we focus in our evaluation on image appeal.

\begin{figure*}[htb!]
    \centering%
    \includegraphics[height=.16\textheight]{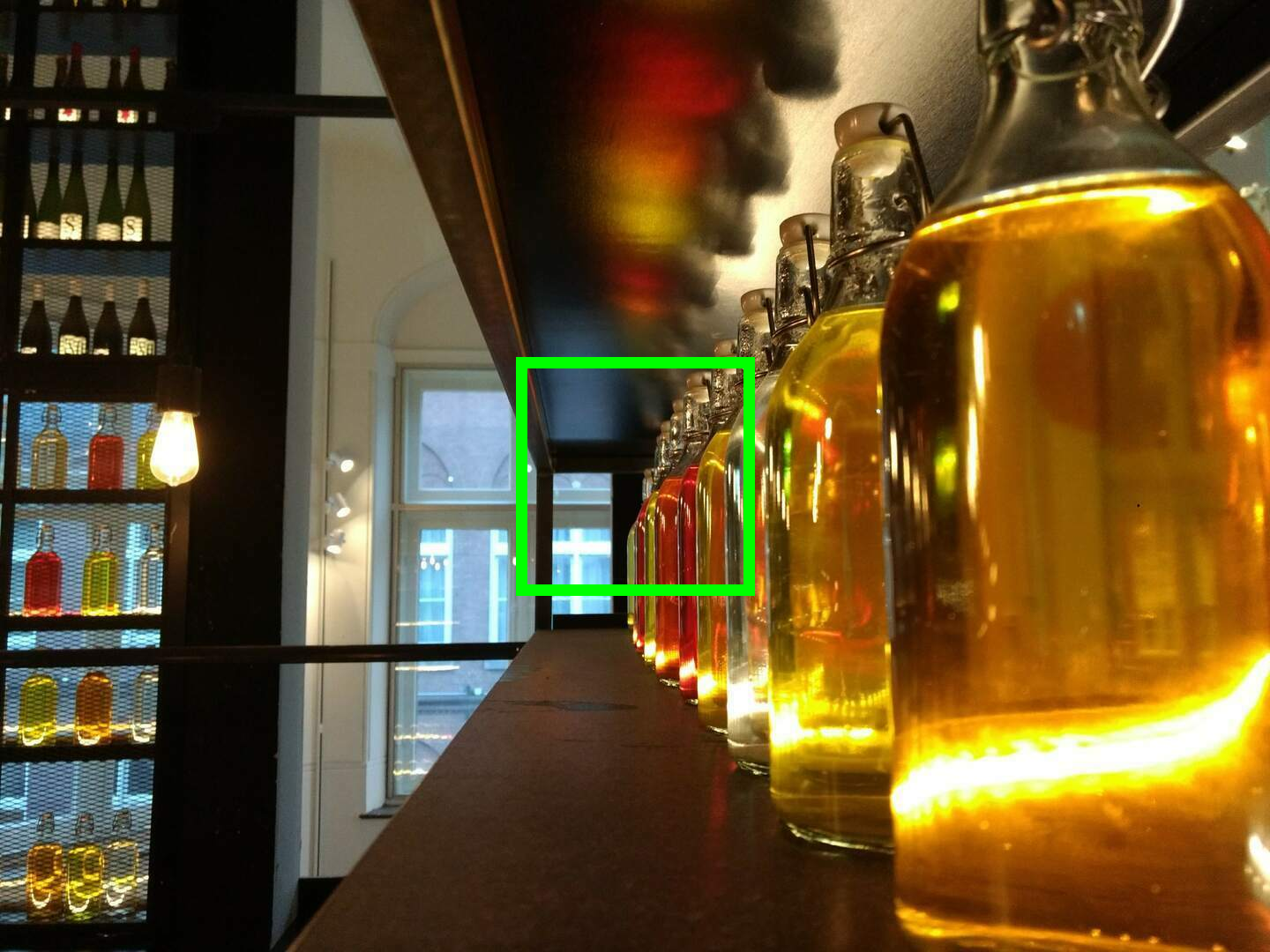} \qquad
    \includegraphics[height=.16\textheight]{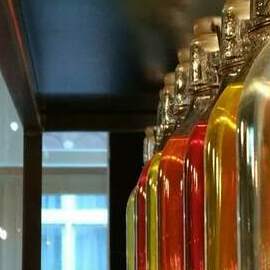}
    \includegraphics[height=.16\textheight]{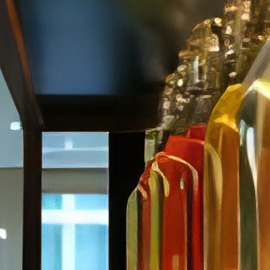}
    \includegraphics[height=.16\textheight]{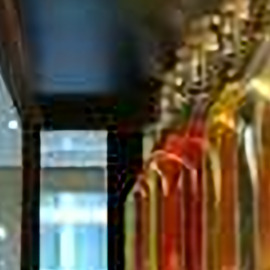}
    \caption{AI-based up-scaling examples; left: full image, then 270x270 pixels center crops are shown of the source image, BSRGAN~\textbf{x4}~\cite{zhang2021designing}, and KXNet~\textbf{x4}~\cite{fu2022kxnet}.\label{fig:example:aiupscaling}}%
\end{figure*}

It has been shown that AI-based image processing, generation, or enhancement introduces new types of artifacts~\cite{goering2023ai,goering2023aiquality}, e.g., as it is also highlighted in Figure~\ref{fig:example:aiupscaling}.
In this example, it is visible that a kind of synthetic look (c.f. BSRGAN \textbf{x4}) of the resulting images or different types of noises or blurriness (compare KXNet \textbf{x4}) are introduced.
Such new distortions require specifically trained and optimized image appeal and quality models.
To train and develop such models, an explicitly annotated dataset is required for a proper evaluation.
In the context of quality assessment and appeal of images, several approaches are possible to gather the required human annotations.
For example, crowd-sourcing has been widely used in various studies before, where it shows promising results~\cite{goeringrao2023crowd,rao2021crowdvideo,goering2023imageappeal,goering2023aiquality,goering2023ai}, e.g., for the evaluation of images and AI-generated content considering appeal or quality.
It is also shown, that such crowd-sourcing tests can be used to gather highly reliable results, which are similar as compared to traditionally conducted lab tests.
Furthermore, it is also possible, with some adjustments, to evaluate high-resolution images or videos with remote testing~\cite{goeringrao2023crowd}, where in contrast, for example, some up-scaling methods are only evaluated using a few low-resolution images (e.g. BSRDM~\cite{yue2022blind}).

In the following paper, we focus on five up-scaling methods and design an online study for the evaluation.
We selected recently published DNN-based up-scaling methods namely BSRGAN~\cite{zhang2021designing}, KXNet~\cite{fu2022kxnet}, Real-ESRGAN~\cite{wang2021realesrgan}, and waifu2x~\cite{waifu2x}, and we also included a Lanczos filter up-scaling approach.
Overall for all of the methods included, we consider two up-scaling factors, namely times two and four, in our evaluation.
Our resulting dataset consists of $1496$ images with human appeal ratings, it covers all up-scaling methods with the two settings and all original images.
The used source images originate from our own image subset (``own'') of the published AVT-ImageAppeal-Dataset~\cite{goering2023imageappeal}.
We re-scaled them to a common height of 1080p, the reason for this is that the display devices in crowd settings are usually restricted and also because 1080p is a commonly used resolution for various devices, such as smartphones, TVs, and computer monitors.

In addition to the evaluation of the visual quality of these newer up-scaling algorithms, it may also be required to understand which method is best for which scenario considering image appeal and which algorithm has been used to upscale an image.
We demonstrated that DNNs can detect which methods have been used for up-scaling.
Furthermore, we propose a model using DNNs to predict the image appeal rating of the generated images.
Our trained model can predict the appeal of the images with a Pearson Correlation of $\approx 0.84$ and can be used to evaluate such AI-based up-scaling algorithms.
In addition, we evaluate state-of-the-art image features and quality models, where we figure out that none of the features or models, or combination of features are able to outperform our transfer-learned DNN model.
The data, trained models, and evaluation code are publicly shared for reproducibility in the context of open science\footnote{\url{https://github.com/Telecommunication-Telemedia-Assessment/ai_upscaling}}.

The remainder of the paper is organized as follows.
In the following Section~\ref{sec:soa} we describe the state-of-the-art for image up-scaling.
Subsequently, in Section~\ref{sec:dataset} the dataset, and how it has been compiled is described.
In Section~\ref{sec::eval} the conducted online study is evaluated and the results are then used in Section~\ref{sec::model} for an automatic approach to predict which up-scaling method is best for a given image.
The paper is summarized in Section~\ref{sec::conclusion} with open points and possible future work.

\section{Related Work} \label{sec:soa}
Image up-scaling is a well-analyzed research field~\cite{singh2020survey,van2006image}, with Nearest-neighbor interpolation being one of the simplest approaches or other signal-based methods that are popular, as for example the Lanczos filter~\cite{duchon1979lanczos}.
The general problem is to provide a high-resolution image based on a low-resolution input image, considering e.g. the details and content of the image.
In the following, we focus on recently published DNN-based approaches for image up-scaling.

In general, the recent advances in deep learning showed promising results of using DNNs for various image processing tasks, where up-scaling is just one example.
For the problem of image up-scaling, also handled as super-resolution, such deep learning-based approaches can be categorized into CNN-based, RNN-CNN-based, and GAN-based models~\cite{ha2018deep} or even more fine granular~\cite{anwar2020deep}.
CNN-based models are, e.g., SRCNN~\cite{dong2014learning}, waifu2x~\cite{waifu2x}, neural-enhance~\cite{shi2016real} or VDSR~\cite{kim2016accurate}, here the low-resolution image is usually extended by features learned from a convolutional neural network and then fused together to form the up-scaled version.
The RNN-CNN-based approaches, such as MemNet~\cite{tai2017memnet}, use recurrent neural networks with memory to enhance the down-scaled image.
The last model group is GAN-based models, for example, models of this type are SRGAN~\cite{ledig2017photo}, Real-ESRGAN~\cite{wang2021realesrgan}, or BSRGAN~\cite{zhang2021designing}.
In general GAN-based models consist of two networks, a generative network and a discriminative network, where the generative network upsamples the image in case of super-resolution and the discriminative network is used to distinguish the ground truth and up-scaled images.
It is, e.g., shown that SRGAN~\cite{ledig2017photo} outperforms signal-based and CNN/RNN-CNN-based approaches.
Real-ESRGAN~\cite{wang2021realesrgan} has been trained on real-world content and also showed promising results compared to other models.
There are also models with mixed approaches, such as KXNet~\cite{fu2022kxnet}, or VDVAE-SR~\cite{chira2023image} available.
Furthermore, not all models, e.g., BSRDM~\cite{yue2022blind} are applicable for higher resolution input images with their provided open-source implementation, mainly because they are just trained and evaluated on smaller inputs as a proof-of-concept.
In addition, some proposed methods are just evaluated with low-resolution input images due to performance reasons.

In most cases, the evaluation focuses primarily on objective metrics such as SSIM, PSNR, or a few example images for demonstration purposes, which is for example the case for KXNet~\cite{fu2022kxnet}, BSRGAN~\cite{zhang2021designing}, or Real-ESRGAN~\cite{wang2021realesrgan}.
However, the complexity and generation of content using DNNs introduce new types of distortions, that the used image quality models cannot handle properly, because they have been developed with different distortions.
For pure performance reasons, it is clear, that PSNR or SSIM can be used for the evaluation to compare the generated image with the high-resolution version.
However, some of the models may still perform reasonably well even though they do not match perfectly the high-resolution reference image, considering that the introduced new content or artifacts may look appealing or real.
Objective quality models need re-training for such specific generated contents, as it is also shown for AI-generated images in~\cite{goering2023aiquality}.
More recent no-reference image quality models may be applicable to evaluate the quality, e.g., DBCNN~\cite{zhang2020blind}, HYPERIQA~\cite{su2020blindly}, Deimeq~\cite{goering2018Deimeq}, or NIMA~\cite{idealods2018imagequalityassessment}.
These image quality models are mostly based on DNNs combined with transfer learning and outperform for image compression and common distortions other state-of-the-art models.
Thus they may be able to handle and detect distortions of AI-generated content which will be covered in our evaluation.

To sum up, we identified three open points which we will be addressing in the following paper.
Firstly, there are only a few studies focusing on the comparison of several AI-based upscaling methods available.
Secondly, high resolution images are less often used for the evaluation.
And thirdly, the evaluation does usually not include human annotations for a wide range of images and modern objective models for AI-based artifacts.

\section{Dataset} \label{sec:dataset}
\begin{figure*}[bt!]
    \centering%
    \includegraphics[width=.99\textwidth]{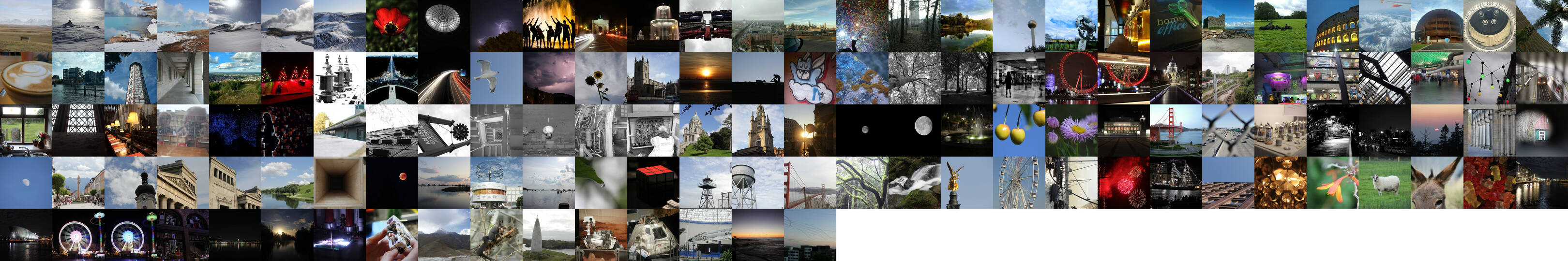}
    \caption{Center crop thumbnails of all used source image for the subjective evaluation; based on the ``own'' subset of AVT-ImageAppeal-Dataset~\cite{goering2023imageappeal}.}
    \label{fig:db:thumbs}%
\end{figure*}

\begin{figure}[bt!]
    \centering%
    \includegraphics[width=.99\columnwidth]{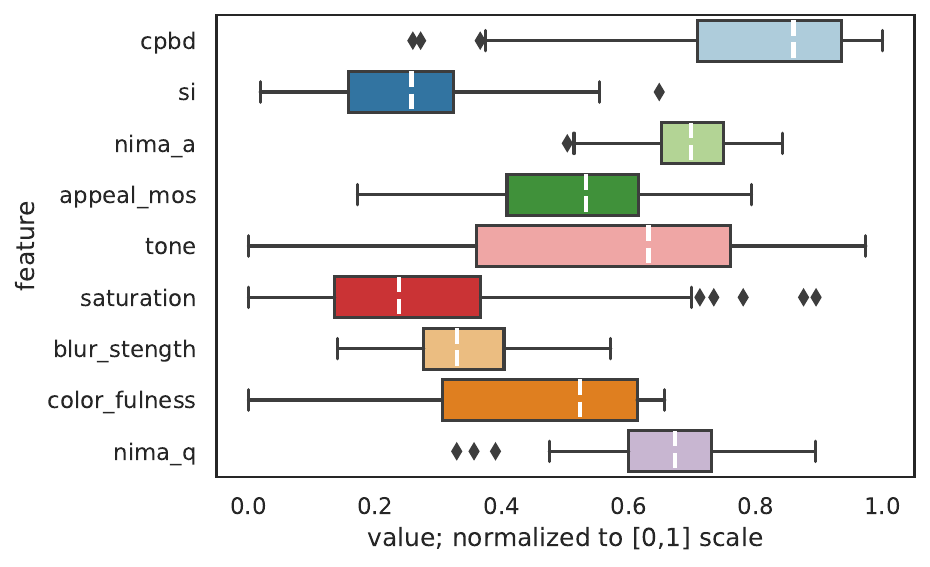}
    \caption{Boxplots of several features calculated for the source images; based on the feature values provided in the AVT-ImageAppeal-Dataset~\cite{goering2023imageappeal}.}
    \label{fig:db:features}%
\end{figure}

To create the dataset, we used four recently published or established DNN-based approaches for image up-scaling, namely, BSRGAN~\cite{zhang2021designing}, KXNet~\cite{fu2022kxnet}, Real-ESRGAN~\cite{wang2021realesrgan}, and waifu2x~\cite{waifu2x}.
In addition to the DNN-based methods, we up-scaled the images using ImageMagick\footnote{\url{https://imagemagick.org}} (version 6.9.10-23) with the Lanczos filter.
The Lanczos up-scaling algorithm is included in our dataset to compare the newly developed methods with traditional signal-based algorithms.
The selection was based on recent models with available and usable open-source implementations, published trained models, which are trained for real photo content, and are applicable for higher resolution output images (1080p).
The used implementations are listed in Table~\ref{tab:upscalingalgos}.
In total, we used 136 images from our published AVT-ImageAppeal-Dataset~\cite{goering2023imageappeal}, these 136 images are all from the ``own'' subset and cover a wide range of realistic images.
The thumbnails of the included images are shown in Figure~\ref{fig:db:thumbs}.
A general characterization considering different features, namely
CPBD~\cite{narvekar2011no}, spatial information (SI)~\cite{recommendation2008p}, nima appeal (nima\_a)~\cite{idealods2018imagequalityassessment}, subjective appeal ratings (appeal\_mos)~\cite{goering2023imageappeal}, tone~\cite{aydin2014automated}, saturation~\cite{aydin2014automated},
blur strength~\cite{crete2007blur}, colorfulness~\cite{hasler2003measuring} and nima quality (nima\_q)~\cite{idealods2018imagequalityassessment}, is shown in Figure~\ref{fig:db:features}.
These features are provided in the published dataset~\cite{goering2023imageappeal} and are re-scaled to [0,1]-values in the Figure, considering their  range of values.
It is visible, that the images of the ``own'' subset form a representative selection of real-world contents.

As a unification step, we re-scaled all images to a common height of 1080\,pixels.
Afterward, we scaled them down to a height of 540\,pixels (540p) and 270\,pixels (270p), respectively.
The 540p images are later up-scaled with a factor of two~(\textbf{x2}), and the 270p images are up-scaled with a factor of four (\textbf{x4}).
For all selected up-scaling methods we perform both up-scaling factors.
The used implementation for Real-ESRGAN did not include a proper up-scaling with \textbf{x2}, we used the model realesrgan-x4plus, and therefore we up-scaled using the factor four and then down-scaled by two.
Using the mentioned approach we have a total number of $136\times(2\times5 + 1)=1496$ images, we included also the 1080p original source images in the subjective evaluation as a hidden reference (labeled as \textbf{x1}).

\begin{table}[htb!]
\scriptsize
\caption{Upscaling methods, their implementations and sources.}
\label{tab:upscalingalgos}
\begin{tabular}{lrr}
    \toprule
    Upscaler    & Implementation                                     & Source \\
    \midrule
    BSRGAN      & \url{https://github.com/cszn/BSRGAN}               & \cite{zhang2021designing} \\
    KXNet       & \url{https://github.com/jiahong-fu/KXNet}          & \cite{fu2022kxnet} \\
    Real-ESRGAN & \url{https://github.com/xinntao/Real-ESRGAN}       & \cite{wang2021realesrgan} \\
    waifu2x     & \url{https://github.com/nihui/waifu2x-ncnn-vulkan} & \cite{waifu2x} \\
    Lanczos     & \url{https://imagemagick.org}                      & \cite{duchon1979lanczos} \\
    \bottomrule
\end{tabular}
\end{table}

\section{Evaluation} \label{sec::eval}
Using the created dataset we designed an online crowd-sourcing test for the evaluation of image appeal.
For the deployment of the online test, we use our publicly available tool AVRate Voyager~\cite{goering2021voyager} with minor adjustments to the provided templates.
AVRate Voyager is a framework to perform audiovisual quality assessment tests in online studies and it has been already used in other works~\cite{goering2023aiquality,goeringrao2023crowd,goering2023imageappeal,goering2023aiquality}.
We modified the rating template, and introduction, and changed the question which is asked to the participants for each stimulus (``Please rate the appeal of the shown image.'').
Furthermore, because it is not feasible for participants to rate all $1496$ images in an online test, we used a similar randomly-based sub-selection approach as described and used in~\cite{goeringrao2023crowd}.
So we asked participants to rate $400$ images out of a total number of $1496$, which resulted in approximately 30 minutes of time required for the subjective test.
We did not include a dedicated training phase to reduce the overall duration of the test.
In total 55 participants took part in the study, excluding incomplete runs of participants.
All participants have been recruited from the Clickworkers platform.
Overall, the images have been rated at least by 4, at most by 25, and on average by 14.7 participants.

\begin{figure}[htb!]
    \centering%
    \includegraphics[width=.99\columnwidth]{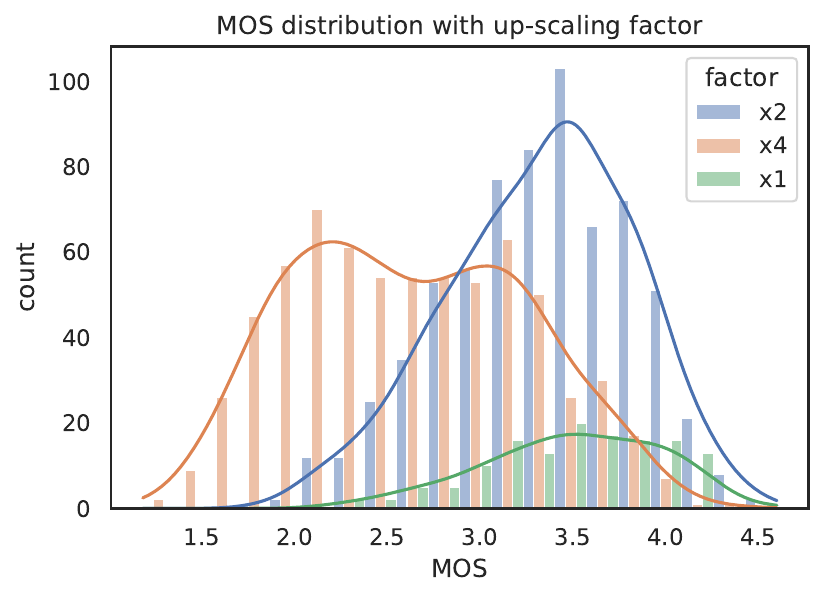}
    \caption{MOS distribution with up-scaling factor, \textbf{x1} refers to the source images with no up-scaling.}
    \label{fig:mos:dist}%
\end{figure}

In Figure~\ref{fig:mos:dist} the MOS distributions for the different up-scaling factors are shown.
It can be seen that the source images \textbf{x1} have the highest MOS values, followed by the up-scaling factor \textbf{x2}.
The factor \textbf{x4} has the lowest scores, here it must be considered that up-scaling with \textbf{x4} is the hardest for the algorithms.
We further performed an SOS analysis~\cite{hossfeld2011sos}, and estimated an $a$ value of $\approx0.275$, which is similar to reported values for such online tests~\cite{goeringrao2023crowd,rao2021crowdvideo,goering2023imageappeal,goering2023ai}.

\begin{figure}[htb!]
    \centering%
    \includegraphics[width=.9\columnwidth]{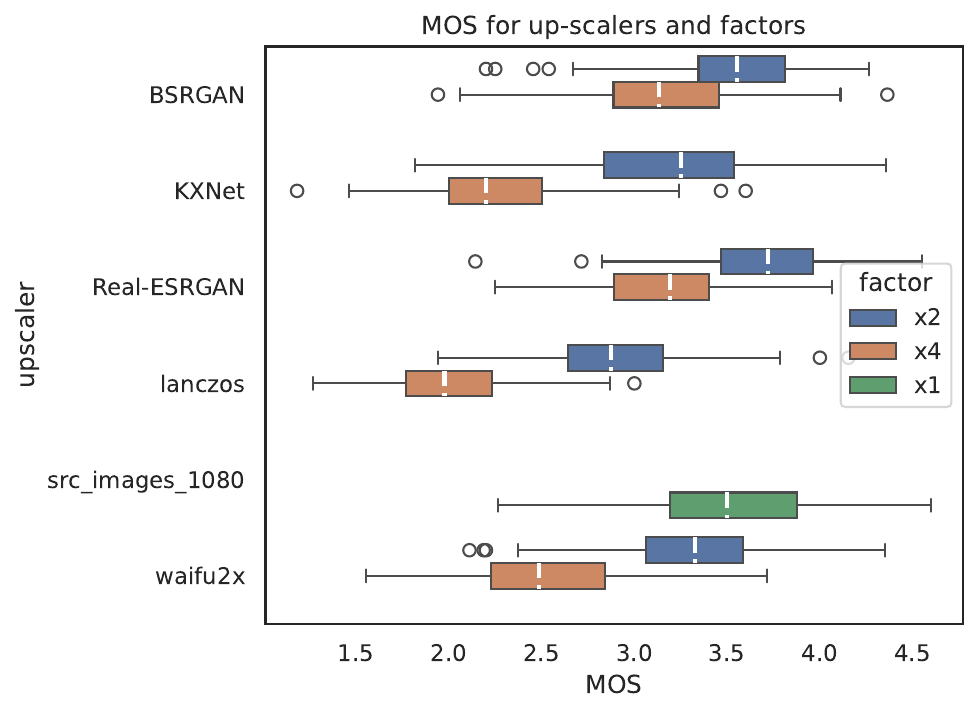}
    \caption{Boxplots for MOS distribution for the up-scaling algorithms.}
    \label{fig:mos:upscalers}%
\end{figure}

\begin{figure}[htb!]
    \centering%
    \includegraphics[width=.9\columnwidth]{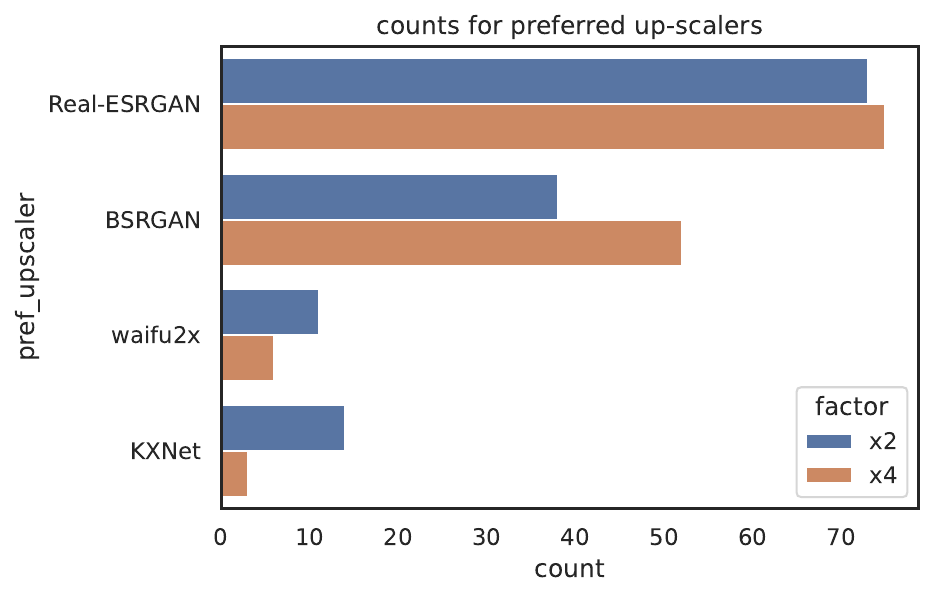}
    \caption{Preference of up-scaling algorithms per up-scaling factor.}
    \label{fig:pref:upscalers}%
\end{figure}

Considering the individual up-scaling methods, compare Figure~\ref{fig:mos:upscalers}, we can conclude that the Real-ESRGAN is the best for both factors, followed by BSRGAN.
The worst appealing methods are Lanczos and KXNet.
The waifu2x method is in between the best and worst methods.
Based on the highest appeal rating for a given up-scaling method, we estimated which method is preferred on a per image basis, shown in Figure~\ref{fig:pref:upscalers}.
The overall results are similar as compared to the overall MOS distribution.
Here, Lanczos has not been preferred for any of the included images.

To further evaluate, whether these methods may change the appeal even beyond the source image appeal, we calculated differences between source image MOS and up-scaled MOS.
Using the differences we defined when this is if $mos(src)- mos(upscaled) > 0$ then the source image is preferred otherwise the upscaled version.
In Table~\ref{tbl:srcprefcount} we counted for each factor this preference.
For the up-scaling factor \textbf{x2} only 35 source images (26\%) have a better appeal than the up-scaled version.
In contrast to the factor \textbf{x4}, where the majority (60\%) preferred the source image.

\begin{table}[htb!]
    \centering
    \scriptsize
    \caption{Source image preferred over up-scaled version}
    \label{tbl:srcprefcount}
    \begin{tabular}{rlrr}
    \toprule
    source image preferred & up-scaling factor & count  & \% \\
    \midrule
    no                     & x2               & 101      & 74 \\
    no                     & x4               & 55       & 40 \\
    \midrule
    yes                    & x2               & 35       & 26 \\
    yes                    & x4               & 81       & 60 \\
    \bottomrule
    \end{tabular}
\end{table}

\begin{figure*}[bth!]
    \centering%
    \includegraphics[height=.16\textheight]{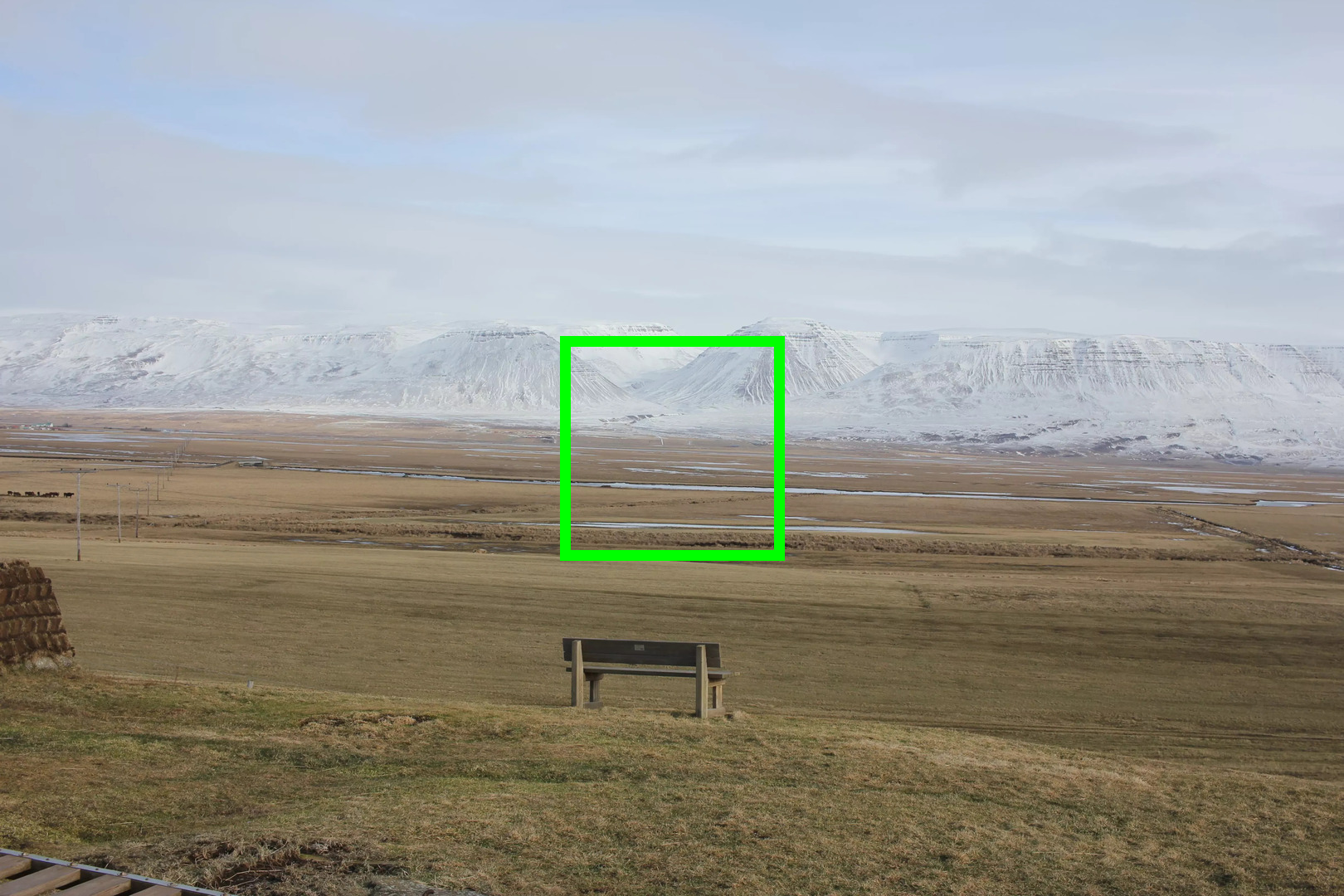} \qquad
    \includegraphics[height=.16\textheight]{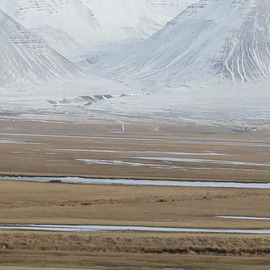}
    \includegraphics[height=.16\textheight]{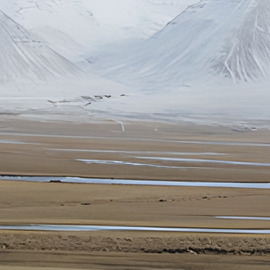}
    \includegraphics[height=.16\textheight]{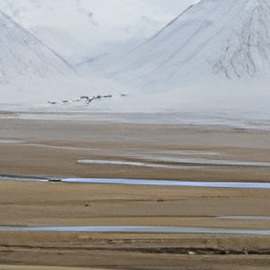}
    \caption{Preference example; left: full source image; then 270x270 pixels center crops are shown of the source image, Real-ESRGAN~\textbf{x2}, and Real-ESRGAN~\textbf{x4}.}
    \label{fig:pref:upscalers:example}%
\end{figure*}

For instance, in Figure~\ref{fig:pref:upscalers:example}, one example image and the corresponding up-scaling variants are shown.
The source image got a mean appeal rating of $\approx 3.78$, the \textbf{x2} Real-ESRGAN variant $\approx 4.0$ and the \textbf{x4} Real-ESRGAN $\approx 3.17$ respectively.
Real-ESRGAN was in both up-scaling cases the best-rated up-scaling algorithm.
So in this case, for up-scaling with \textbf{x2} the Real-ESRGAN is preferred over the source image, in contrast to an up-scaling with \textbf{x4} where the source image is preferred by the appeal rating, due to the lack of details, smoothed and artificial look of the resulting images.

\section{Detection} \label{sec::detection}
To evaluate which up-scaling method has been used, we trained various DNNs in a multi-class classification setup similar to the approach used in \cite{goering2021rules,goering2023ruleext}.
We use a pre-trained baseline DNN, which has been trained for ImageNet~\cite{ILSVRC15}, and perform transfer-learning~\cite{torrey2010transfer}.
Because our images are high-resolution images, and considering that we want to detect the up-scaling method, we split each 1080p image in several patches each with a size of 224x224 (with no overlap).
This size was selected because it is the input size of the majority of the considered baseline DNNs.
Afterwards a baseline-specific pre-processing is performed within the network.
The split approach of the high-resolution images resulted in $40.832$ patches, which are used for training and validation.
For the baseline DNN, we removed the last classification-specific layer and added a flattening layer.
This is followed by a dropout layer with a rate of $0.2$, a dense fully connected layer with $n$ output signals ($n=6$ for the five up-scaling methods and the source images), and softmax activation.
The flattening, dropout, and dense layers are specific to our classification task and are the parameters to be trained.
The approach allows the usage of several pre-trained baseline DNNs that are used as they are, thus no re-training for the baseline models is performed.
The implementation is based on Keras~\cite{chollet2015keras} using Python~3.9.

\begin{figure}[htb!]
    \centering%
    \includegraphics[width=.99\columnwidth]{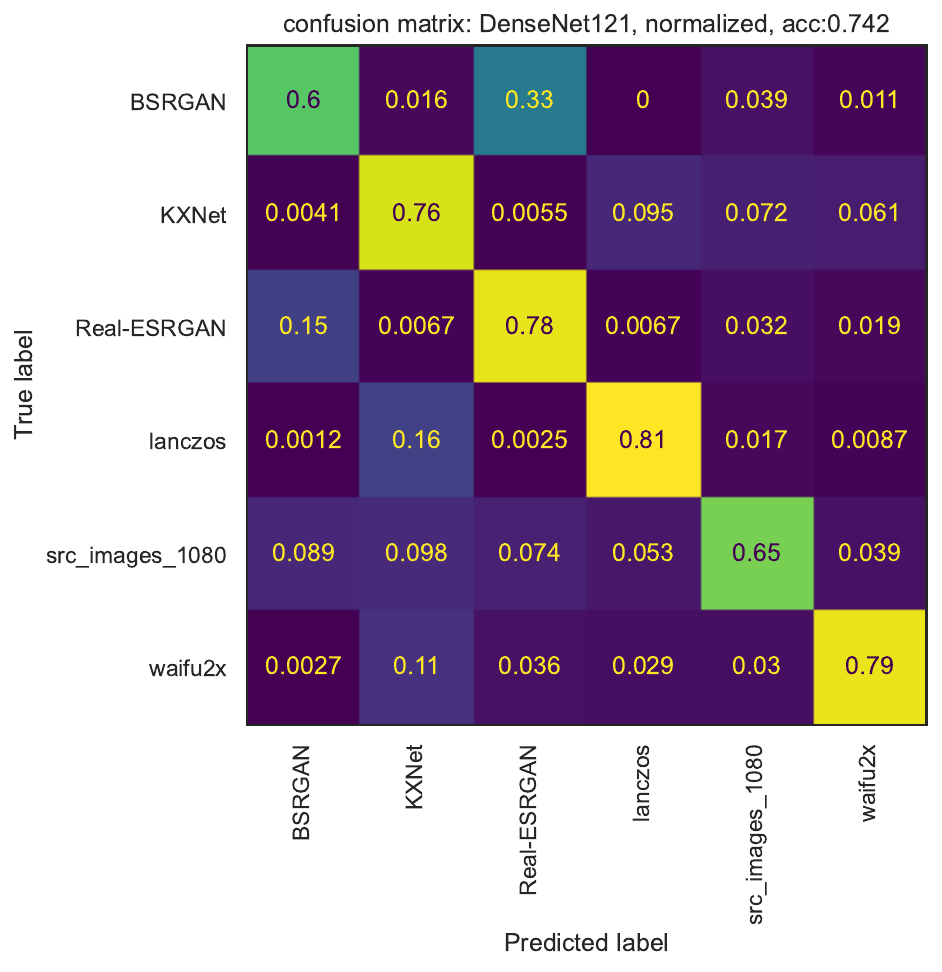}
    \caption{Confusion matrix of the best model (DenseNet121) for the detection of the used up-scaling method.}
    \label{fig:detection:confusion}%
\end{figure}

In total, we trained and evaluated 16 models, using a 90\%-10\% train-validation splitting of the dataset.
The performance values, considering accuracy, f1-score, precision, recall and Matthew's correlation coefficient (mcc) for all models are listed in Table~\ref{tbl:detection}.
Best performing models are DenseNet or ResNet model variants, worst performing are Inception and VGG based models.
\begin{table}[bht!]
    \centering
    \scriptsize
    \caption{Performance values of all DNN models for the detection of the used up-scaling method; rounded to 3 decimals.}
    \label{tbl:detection}
    \begin{tabular}{p{1.8cm}rrrrr}
    \toprule
    Model             & Accuracy & f1-score & Precision & Recall & mcc \\
    \midrule
    DenseNet121       & 0.742    & 0.743    & 0.750     & 0.742  & 0.689\\
    DenseNet169       & 0.735    & 0.735    & 0.743     & 0.735  & 0.680\\
    ResNet50          & 0.725    & 0.724    & 0.723     & 0.725  & 0.666\\
    DenseNet201       & 0.724    & 0.723    & 0.725     & 0.724  & 0.666\\
    ResNet152         & 0.701    & 0.699    & 0.706     & 0.701  & 0.639\\
    ResNet101         & 0.700    & 0.697    & 0.697     & 0.700  & 0.636\\
    MobileNetV2       & 0.698    & 0.698    & 0.699     & 0.698  & 0.634\\
    MobileNet         & 0.685    & 0.683    & 0.684     & 0.685  & 0.618\\
    Xception          & 0.566    & 0.565    & 0.566     & 0.566  & 0.473\\
    ResNet50V2        & 0.559    & 0.558    & 0.558     & 0.559  & 0.465\\
    ResNet101V2       & 0.543    & 0.539    & 0.537     & 0.543  & 0.445\\
    ResNet152V2       & 0.524    & 0.523    & 0.523     & 0.524  & 0.422\\
    % NASNetMobile    & 0.521    & 0.523    & 0.539     & 0.521  & 0.421\\  %excluded because it is not part in the other experiment
    VGG16             & 0.520    & 0.520    & 0.523     & 0.520  & 0.417\\
    InceptionV3       & 0.494    & 0.492    & 0.515     & 0.494  & 0.388\\
    InceptionResNetV2 & 0.484    & 0.484    & 0.491     & 0.484  & 0.375\\
    VGG19             & 0.479    & 0.475    & 0.478     & 0.479  & 0.367\\
    \bottomrule
    \end{tabular}
\end{table}
The best performing model was DenseNet121, with an accuracy of $\approx0.74$, f1-score $\approx0.74$, precision $\approx0.75$ and recall $\approx0.74$ for the validation data.
In Figure~\ref{fig:detection:confusion} the confusion matrix for the DenseNet121 model is shown.

In general, this evaluation shows that it is possible to detect which up-scaling method has been used, assuming all algorithms are known beforehand.
The approach is also to be seen as a proof-of-concept.
For a better detection approach more images and more algorithms would be required.

\section{Appeal Prediction} \label{sec::model}
Similarly, as compared to the detection prediction, we use transfer learning, with in total 16 DNNs.
Instead of the classification layer as the last layer, we added a dense layer with 1024 signals and ReLu activation and an additional dense layer with one output signal and softmax activation for the final prediction.
The mean appeal scores have been normalized to $[0,1]$.
We used a 90\%-10\% train-validation split, and center cropped the images before to 224x224 which is the input size of the DNNs, center cropping has been successfully used before for image appeal and video quality evaluation~\cite{goeringrao2023crowd,goering2019cencro}.
The performance values of all models are listed in Table~\ref{tbl:appealpred}.

\begin{table}[htb!]
    \centering
    \scriptsize
    \caption{Performance values of all DNN models for image appeal prediction; DenseNet~\cite{huang2017densely} and ResNet~\cite{he2016deep} variants best.}
    \label{tbl:appealpred}
    \begin{tabular}{lrrr}
    \toprule
    Model & Pearson & Kendall & Spearman \\
    \midrule
      ResNet152V2 &  0.838 &  0.622 &  0.811 \\
      DenseNet121 &  0.823 &  0.611 &  0.801 \\
      DenseNet169 &  0.821 &  0.598 &  0.787 \\
      ResNet101V2 &  0.818 &  0.601 &  0.794 \\
      DenseNet201 &  0.817 &  0.595 &  0.786 \\
       ResNet50V2 &  0.816 &  0.598 &  0.791 \\
         Xception &  0.814 &  0.581 &  0.767 \\
        MobileNet &  0.807 &  0.586 &  0.780 \\
InceptionResNetV2 &  0.742 &  0.527 &  0.717 \\
            VGG16 &  0.708 &  0.487 &  0.676 \\
            VGG19 &  0.671 &  0.470 &  0.661 \\
         ResNet50 &  0.345 &  0.253 &  0.369 \\
        ResNet101 &  0.319 &  0.265 &  0.386 \\
        ResNet152 &  0.207 &  0.213 &  0.318 \\
      MobileNetV2 &  0.052 &  0.030 &  0.044 \\
      InceptionV3 & -0.067 & -0.023 & -0.028 \\
    \bottomrule
    \end{tabular}
\end{table}

\begin{figure}[tb!]
    \centering%
    \includegraphics[width=.9\columnwidth]{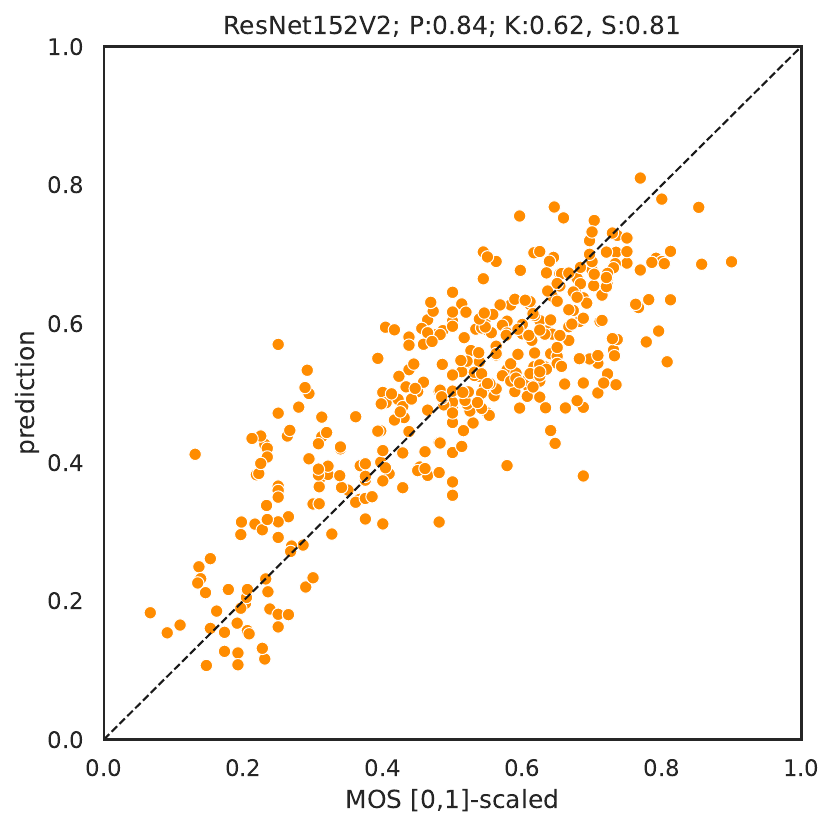}
    \caption{Scatter plot of the best performing image appeal prediction model (ResNet152V2).}
    \vspace{-1em}
    \label{fig:appeal:best}%
\end{figure}

The results indicate that appeal prediction for the considered up-scaling algorithms is possible, however, it should be mentioned that the overall number of images may be too limited for the general applicability of the model.
Thus the numbers are just an indicator of whether it is possible, which is confirmed, see similar~\cite{goering2023imageappeal}.
The best-performing model is ResNet152V2, compare Figure~\ref{fig:appeal:best}, followed by DenseNet121 with nearly similar performance.
Only MobileNetV2 and InceptionV3 have worse performance for this task, this may be due to the architecture or age of these models, as it is also shown in~\cite{goering2018Deimeq} for image quality or for image appeal~\cite{goering2023imageappeal}.

Furthermore, we calculated 11 features for image appeal and quality, which are part of the open source code from~\cite{goering2023ai}.
In Table~\ref{fig:features} the results are summarized.
The \textbf{*combined*} variant refers to a 10-fold cross-validation run of a Random Forest Model (RF), with 100 trees and default parameters of scikit-learn~\cite{scikit-learn}.

\begin{table}[tb!]
    \centering
    \scriptsize
    \caption{Image features compared to image appeal rating; values rounded to 3 decimals, sorted by Pearson.}
    \label{fig:features}
    \begin{tabular}{lrrr}
    \toprule
    Feature               & Pearson & Kendall & Spearman \\
            \midrule

    \textbf{*combined*} & 0.669   & 0.450   & 0.631 \\
             \midrule
    cpbd                  & 0.572   & 0.363   & 0.522 \\
    fft                   & 0.331   & 0.238   & 0.350 \\
    noise                 & 0.299   & 0.302   & 0.447 \\
    si                    & 0.213   & 0.135   & 0.200 \\
    blur                  & 0.152   & 0.110   & 0.164 \\
    saturation            & 0.093   & 0.057   & 0.087 \\
    colorfulness          & 0.024   & 0.019   & 0.028 \\
    contrast              & -0.013  & -0.005  & -0.008 \\
    tone                  & -0.039  & -0.020  & -0.031 \\
    niqe                  & -0.088  & -0.040  & -0.061 \\
    blur strength         & -0.378  & -0.257  & -0.375 \\
    \bottomrule
    \end{tabular}
\end{table}
It is visible in Table~\ref{fig:features} that none of the signal features has a strong correlation with the appeal ratings, only CPBD~\cite{narvekar2011no}, FFT~\cite{goering2021pixel} and noise~\cite{goering2021pixel} have a medium correlation.
The combined model is better than the individual features, however, it has still a lower performance than the re-trained DNN model.

Furthermore, other state-of-the-art no-reference image quality models have been included in the evaluation, here, the implementation ``IQA-PyTorch'' provided by~\citeauthor{pyiqa}~\cite{pyiqa} has been used.
In Table~\ref{fig:iqa:selected}, an overview of the results for selected models is given, only models with at least a medium Pearson correlation are included, and all scores for all models are included in the shared data.
The best-performing models are DBCNN~\cite{zhang2020blind}, and HYPERIQA~\cite{su2020blindly} with a medium Pearson correlation, which is comparable to the \textbf{*combined*} model shown in Table~\ref{fig:features} and is still worse than the re-trained DNNs.
We further checked the NIMA model~\cite{idealods2018imagequalityassessment}, however, due to the fact that it down-scales the image for processing it cannot be used for up-scaling quality or appeal evaluation.
Therefore we used a 224x224 pixels center crop of the images and added this to the table as ``NIMA quality CC'', which also did improve the performance.
We furthermore include some full-reference image quality models using the PyTorch Image Quality (PIQ) Toolbox~\cite{kastryulin2022piq}, namely MS SSIM, VIF, PSNR, and SSIM.
Other models of this toolbox had a lower overall performance, and are not included in the table.
None of the full-reference models has a good performance considering Pearson Correlation, this may be due to the new and unknown introduced artificial artifacts while up-scaling.

\begin{table}[tb!]
    \centering
    \scriptsize
    \caption{Selected image quality models compared to image appeal rating; values rounded to 3 decimals, sorted by Pearson.}
    \label{fig:iqa:selected}
    \begin{tabular}{lrrr}
    \toprule
    Model           & Pearson & Kendall & Spearman \\
    \midrule
    DBCNN           & 0.605   & 0.436   & 0.618 \\
    HYPERIQA        & 0.601   & 0.414   & 0.592 \\
    CNNIQA          & 0.592   & 0.376   & 0.536 \\
    MUSIQ           & 0.555   & 0.365   & 0.522 \\
    MANIQA          & 0.505   & 0.345   & 0.493 \\
    paq2piq         & 0.492   & 0.311   & 0.448 \\
    NIMA quality CC & 0.433   & 0.281   & 0.408 \\
    \midrule

    ms\_ssim        & 0.368   & 0.232   & 0.348 \\
    vif             & 0.363   & 0.248   & 0.373 \\
    psnr            & 0.248   & 0.164   & 0.244 \\
    ssim            & 0.183   & 0.135   & 0.203 \\

    % DBCNN~\cite{zhang2020blind}                               & 0.605   & 0.436   & 0.618 \\
    % HYPERIQA~\cite{su2020blindly}                             & 0.601   & 0.414   & 0.592 \\
    % CNNIQA~\cite{kang2014convolutional}                       & 0.592   & 0.376   & 0.536 \\
    % MUSIQ~\cite{ke2021musiq}                                  & 0.555   & 0.365   & 0.522 \\
    % MANIQA~\cite{yang2022maniqa}                              & 0.505   & 0.345   & 0.493 \\
    % paq2piq~\cite{ying2020patches}                            & 0.492   & 0.311   & 0.448 \\
    % NIMA quality CC~\cite{idealods2018imagequalityassessment} & 0.433   & 0.281   & 0.408 \\
    \bottomrule
    \end{tabular}
\end{table}

\section{Conclusion and Future Work} \label{sec::conclusion}
We started with the observation that DNN-based up-scaling algorithms seem to perform better than traditional signal-based approaches.
However, the majority of studies do not compare several of such algorithms and usually do not include a large-scale subjective evaluation.
For this reason, we selected five different open-source algorithms, namely Real-ESRGAN, KXNet, BSRGAN, waifu2x, and Lanczos, and created a dataset considering two up-scaling factors (\textbf{x2} and \textbf{x4}).
This dataset consists of 136 base images, a subset from the AVT-ImageAppeal-Dataset~\cite{goering2023imageappeal}, which have been up-scaled to $1496$ variants using the algorithms and further extended by human annotations considering image appeal.
We selected image appeal as an evaluation criterion, due to a high similarity to image quality and also because such new algorithms may introduce artifacts that are artificial.
To gather the annotations we carried out a crowd-sourcing test, and showed in the evaluation that the most appealing model in most cases is Real-ESRGAN.
The second best model is BSRGAN.
While Lanczos has not been preferred considering image appeal for any included image and up-scaling factor.
In addition to this, we used the images to train a detection DNN to classify which up-scaling method has been used, the approach shows promising results.
The best-performing DNN for the detection is DenseNet121.
Furthermore, image appeal can be also predicted automated using DNNs and partially with state-of-the-art models.
Here, our transfer-learned DNN outperforms other models, the model model was ResNet152V2.
However, also signal-based features can be used to predict image appeal in this context, as we show by training a random forest regression model using various extracted signal features.
Overall, it can be stated that AI-based up-scaling algorithms introduce new distortions, which image quality or appeal models must be adjusted to.
Here, more tests with a larger number of source images considering more up-scaling methods are required for the training and development of quality and appeal prediction models.
In future work, newer ai-based up-scaling methods could be used for video up-scaling, here the temporal coherence of the generated images is also an important aspect, which needs to be analyzed.

\section*{Acknowledgment}

The authors would like to thank the participants for taking part in this crowd test.
Furthermore, we want to thank the ``AG Wissenschaftliches Rechnen'' of the TU Ilmenau for providing computing resources.

%\clearpage
\section*{References}
{
\begingroup
    %\setstretch{0.8}
    \renewcommand{\bibfont}{\small}
    %\setlength\bibitemsep{0pt}
    %\sloppy
    \printbibliography[heading=none]

@INPROCEEDINGS{goering2023aiquality,
  author={Steve {G{\"o}ring} and Rakesh {Rao Ramachandra Rao} and Alexander Raake},
  title="Appeal and quality assessment for AI-generated images",
  booktitle="15th International Conference on Quality of Multimedia Experience (QoMEX)",
  year={2023},
}

@article{goering2023imageappeal,
  title={Image Appeal Revisited: Analysis, new Dataset and Prediction Models},
  author={Steve G\"oring and Alexander Raake},
  journal={IEEE Access},
  year={2023},
  publisher={IEEE},
  volume={11},
  number={},
  pages={69563-69585},
  doi={10.1109/ACCESS.2023.3292588},
}

@inproceedings{goering2023ruleext,
  author={Steve {G{\"o}ring} and Rasmus Merten and Alexander Raake},
  title="DNN-based Photography Rule Prediction using Photo Tags",
  booktitle="15th International Conference on Quality of Multimedia Experience (QoMEX)",
  year={2023},
}

@article{goering2023ai,
  title={Analysis of Appeal for realistic AI-generated Photos},
  author={Steve G\"oring and Rakesh {Rao Ramachandra Rao} and Rasmus Merten and Alexander Raake},
  journal={IEEE Access},
  year={2023},
  url={https://doi.org/10.1109/ACCESS.2023.3267968},
  volume={11},
  number={},
  pages={38999-39012},
  doi={10.1109/ACCESS.2023.3267968}
}

@inproceedings{hasler2003measuring,
  title={Measuring colorfulness in natural images},
  author={Hasler, David and Suesstrunk, Sabine E},
  booktitle={Human vision and electronic imaging VIII},
  volume={5007},
  pages={87--95},
  year={2003},
  organization={SPIE}
}

@article{aydin2014automated,
  title={Automated aesthetic analysis of photographic images},
  author={Ayd{\i}n, Tun{\c{c}} Ozan and Smolic, Aljoscha and Gross, Markus},
  journal={IEEE transactions on visualization and computer graphics},
  volume={21},
  number={1},
  pages={31--42},
  year={2014},
  publisher={IEEE}
}

@article{goering2021pixel,
  title={Modular Framework and Instances of Pixel-based Video Quality Models for UHD-1/4K},
  author={Steve G\"oring and Rakesh {Rao Ramachandra Rao} and Bernhard Feiten and Alexander Raake},
  journal={IEEE Access},
  volume={9},
  pages={31842-31864},
  year={2021},
  publisher={IEEE},
  doi={10.1109/ACCESS.2021.3059932},
  url={https://ieeexplore.ieee.org/document/9355144},
  code={https://github.com/Telecommunication-Telemedia-Assessment/pixelmodels}
}

@article{recommendation2008p,
  author       = {Recommendation, ITUT},
  date         = {2008},
  journaltitle = {Int. Telecommunication Union, Tech. Rep},
  title        = {P.910, Subjective video quality assessment methods for multimedia applications,}
}

@article{narvekar2011no,
  title={A no-reference image blur metric based on the cumulative probability of blur detection (CPBD)},
  author={Narvekar, Niranjan D and Karam, Lina J},
  journal={IEEE Transactions on Image Processing},
  volume={20},
  number={9},
  pages={2678--2683},
  year={2011},
  publisher={IEEE}
}

@inproceedings{crete2007blur,
  title={The blur effect: perception and estimation with a new no-reference perceptual blur metric},
  author={Crete, Frederique and Dolmiere, Thierry and Ladret, Patricia and Nicolas, Marina},
  booktitle={Human vision and electronic imaging XII},
  volume={6492},
  pages={196--206},
  year={2007},
  organization={SPIE}
}

@article{scikit-learn,
 title={Scikit-learn: Machine Learning in Python},
 author={Pedregosa, F. and Varoquaux, G. and Gramfort, A. and Michel, V.
         and Thirion, B. and Grisel, O. and Blondel, M. and Prettenhofer, P.
         and Weiss, R. and Dubourg, V. and Vanderplas, J. and Passos, A. and
         Cournapeau, D. and Brucher, M. and Perrot, M. and Duchesnay, E.},
 journal={Journal of Machine Learning Research},
 volume={12},
 pages={2825--2830},
 year={2011}
}

@misc{pyiqa,
  title={{IQA-PyTorch}: PyTorch Toolbox for IQA},
  author={Chaofeng Chen and Jiadi Mo},
  year={2023},
  howpublished = "\url{https://github.com/chaofengc/IQA-PyTorch}"
}

@article{zhang2020blind,
  title={Blind Image Quality Assessment Using A Deep Bilinear Convolutional Neural Network},
  author={Zhang, Weixia and Ma, Kede and Yan, Jia and Deng, Dexiang and Wang, Zhou},
  journal={Transaction on Circuits and Systems for Video Technology},
  volume={30},
  number={1},
  pages={36--47},
  year={2020}
}

@inproceedings{goering2021voyager,
  title={AVRate Voyager: an open source online testing platform},
  author={Steve G\"oring and Rakesh {Rao Ramachandra Rao} and Stephan Fremerey and Alexander Raake},
  year={2021},
  booktitle={23rd International Workshop on Multimedia Signal Processing (MMSP)},
  pages={1--6},
  organization={IEEE}
}

@InProceedings{wang2021realesrgan,
    author    = {Xintao Wang and Liangbin Xie and Chao Dong and Ying Shan},
    title     = {Real-ESRGAN: Training Real-World Blind Super-Resolution with Pure Synthetic Data},
    booktitle={Proceedings of the IEEE/CVF international conference on computer vision},
    date      = {2021}
}

@misc{idealods2018imagequalityassessment,
  title={Image Quality Assessment},
  author={Christopher Lennan and Hao Nguyen and Dat Tran},
  year={2018},
  howpublished={\url{https://github.com/idealo/image-quality-assessment}}
}

@inproceedings{goering2018Deimeq,
  title={deimeq -- A Deep Neural Network Based Hybrid No-reference Image Quality Model},
  author={Steve G{\"{o}}ring and Alexander Raake},
  booktitle={7th European Workshop on Visual Information Processing (EUVIP)},
  pages={1--6},
  year={2018},
  url={https://ieeexplore.ieee.org/document/8611703},
  organization={IEEE}
}

@inproceedings{goering2021rules,
  author={Steve {G{\"o}ring} and Alexander Raake},
  title="Rule of Thirds and Simplicity for Image Aesthetics using Deep Neural Networks",
  year={2021},
  booktitle={23rd International Workshop on Multimedia Signal Processing (MMSP)},
  pages={1--6},
  organization={IEEE}
}

@inproceedings{hossfeld2011sos,
  title={SOS: The MOS is not enough!},
  author={Ho{\ss}feld, Tobias and Schatz, Raimund and Egger, Sebastian},
  booktitle={3rd International Workshop on Quality of Multimedia Experience (QoMEX)},
  pages={131--136},
  year={2011},
  organization={IEEE}
}

@article{goeringrao2023crowd,
  title={Quality assessment of higher resolution images and videos with remote testing},
  author={G{\"o}ring, Steve and Rao, Rakesh Rao Ramachandra and Raake, Alexander},
  journal={Quality and User Experience (QUEX)},
  volume={8},
  number={1},
  pages={2},
  year={2023},
  publisher={Springer}
}

@inproceedings{rao2021crowdvideo,
  author={Rakesh {Rao Ramachandra Rao} and Steve {G{\"o}ring} and Alexander Raake},
  title="Towards High Resolution Video Quality Assessment in the Crowd",
  booktitle="13th International Conference on Quality of Multimedia Experience (QoMEX)",
  pages={1--6},
  year={2021},
  url={https://ieeexplore.ieee.org/document/9465425}
}

@inproceedings{pathak2016context,
  title={Context encoders: Feature learning by inpainting},
  author={Pathak, Deepak and Krahenbuhl, Philipp and Donahue, Jeff and Darrell, Trevor and Efros, Alexei A},
  booktitle={Proceedings of the IEEE conference on computer vision and pattern recognition},
  pages={2536--2544},
  year={2016}
}

@article{salmona2022deoldify,
  title={Deoldify: A review and implementation of an automatic colorization method},
  author={Salmona, Antoine and Bouza, Luc{\'\i}a and Delon, Julie},
  journal={Image Processing On Line},
  volume={12},
  pages={347--368},
  year={2022}
}

@inproceedings{fu2022kxnet,
  title={KXNet: A model-driven deep neural network for blind super-resolution},
  author={Fu, Jiahong and Wang, Hong and Xie, Qi and Zhao, Qian and Meng, Deyu and Xu, Zongben},
  booktitle={European Conference on Computer Vision},
  pages={235--253},
  year={2022},
  organization={Springer}
}

@inproceedings{zhang2021designing,
  title={Designing a practical degradation model for deep blind image super-resolution},
  author={Zhang, Kai and Liang, Jingyun and Van Gool, Luc and Timofte, Radu},
  booktitle={Proceedings of the IEEE/CVF International Conference on Computer Vision},
  pages={4791--4800},
  year={2021}
}

@inproceedings{ronneberger2015u,
  title={U-net: Convolutional networks for biomedical image segmentation},
  author={Ronneberger, Olaf and Fischer, Philipp and Brox, Thomas},
  booktitle={18th International Conference of Medical Image Computing and Computer-Assisted Intervention (MICCAI)},
  pages={234--241},
  year={2015},
  organization={Springer}
}

@inproceedings{wang2018esrgan,
  title={Esrgan: Enhanced super-resolution generative adversarial networks},
  author={Wang, Xintao and Yu, Ke and Wu, Shixiang and Gu, Jinjin and Liu, Yihao and Dong, Chao and Qiao, Yu and Change Loy, Chen},
  booktitle={Proceedings of the European conference on computer vision (ECCV) workshops},
  year={2018}
}

@inproceedings{chira2023image,
  title={Image super-resolution with deep variational autoencoders},
  author={Chira, Darius and Haralampiev, Ilian and Winther, Ole and Dittadi, Andrea and Li{\'e}vin, Valentin},
  booktitle={Computer Vision--ECCV 2022 Workshops: Tel Aviv, Israel, October 23--27, 2022, Proceedings, Part II},
  pages={395--411},
  year={2023},
  organization={Springer}
}

@online{waifu2x,
  author = {nagadomi},
    title = {waifu2x -- \url{https://github.com/nagadomi/waifu2x}}
}

@article{laine2019high,
  title={High-quality self-supervised deep image denoising},
  author={Laine, Samuli and Karras, Tero and Lehtinen, Jaakko and Aila, Timo},
  journal={Advances in Neural Information Processing Systems},
  volume={32},
  year={2019}
}

@article{ILSVRC15,
Author = {Olga Russakovsky and Jia Deng and Hao Su and Jonathan Krause and Sanjeev Satheesh and Sean Ma and Zhiheng Huang and Andrej Karpathy and Aditya Khosla and Michael Bernstein and Alexander C. Berg and Li Fei-Fei},
Title = {{ImageNet Large Scale Visual Recognition Challenge}},
Year = {2015},
journal={International journal of computer vision},
doi = {10.1007/s11263-015-0816-y},
volume={115},
number={3},
pages={211-252}
}

@incollection{torrey2010transfer,
  title={Transfer learning},
  author={Torrey, Lisa and Shavlik, Jude},
  booktitle={Handbook of research on machine learning applications and trends: algorithms, methods, and techniques},
  pages={242--264},
  year={2010},
  publisher={IGI global}
}

@misc{chollet2015keras,
  title={Keras},
  author={Chollet, Fran\c{c}ois and others},
  year={2015},
  howpublished={\url{https://keras.io}}
}

@inproceedings{goering2019cencro,
    author = {Steve G\"oring and Christopher Kr\"ammer and Alexander Raake},
    title = {cencro -- Speedup of Video Quality Calculation using Center Cropping},
    booktitle={21st  International Symposium on Multimedia (ISM)},
    organization={IEEE},
    year = {2019},
    pages={1-8},
    volume={},
    month={12}
}

@inproceedings{su2020blindly,
  title={Blindly assess image quality in the wild guided by a self-adaptive hyper network},
  author={Su, Shaolin and Yan, Qingsen and Zhu, Yu and Zhang, Cheng and Ge, Xin and Sun, Jinqiu and Zhang, Yanning},
  booktitle={Proceedings of the IEEE/CVF conference on computer vision and pattern recognition},
  pages={3667--3676},
  year={2020}
}

@inproceedings{huang2017densely,
  title={Densely connected convolutional networks},
  author={Huang, Gao and Liu, Zhuang and Van Der Maaten, Laurens and Weinberger, Kilian Q},
  booktitle={Proceedings of the IEEE conference on computer vision and pattern recognition},
  pages={4700--4708},
  year={2017}
}

@inproceedings{he2016deep,
  title={Deep residual learning for image recognition},
  author={He, Kaiming and Zhang, Xiangyu and Ren, Shaoqing and Sun, Jian},
  booktitle={Proceedings of the IEEE conference on computer vision and pattern recognition},
  pages={770--778},
  year={2016}
}

@inproceedings{ha2018deep,
  title={Deep learning based single image super-resolution: A survey},
  author={Ha, Viet Khanh and Ren, Jinchang and Xu, Xinying and Zhao, Sophia and Xie, Gang and Vargas, Valentin Masero},
  booktitle={Advances in Brain Inspired Cognitive Systems: 9th International Conference, BICS 2018, Xi'an, China, July 7-8, 2018, Proc. 9},
  pages={106--119},
  year={2018},
  organization={Springer}
}

@article{van2006image,
  title={Image super-resolution survey},
  author={Van Ouwerkerk, JD},
  journal={Image and vision Computing},
  volume={24},
  number={10},
  pages={1039--1052},
  year={2006},
  publisher={Elsevier}
}

@article{singh2020survey,
  title={Survey on single image based super-resolutionimplementation challenges and solutions},
  author={Singh, Amanjot and Singh, Jagroop},
  journal={Multimedia Tools and Applications},
  volume={79},
  pages={1641--1672},
  year={2020},
  publisher={Springer}
}

@article{duchon1979lanczos,
  title={Lanczos filtering in one and two dimensions},
  author={Duchon, Claude E},
  journal={Journal of Applied Meteorology and Climatology},
  volume={18},
  number={8},
  pages={1016--1022},
  year={1979}
}

@inproceedings{dong2014learning,
  title={Learning a deep convolutional network for image super-resolution},
  author={Dong, Chao and Loy, Chen Change and He, Kaiming and Tang, Xiaoou},
  booktitle={Computer Vision},
  pages={184--199},
  year={2014},
  organization={Springer}
}

@inproceedings{kim2016accurate,
  title={Accurate image super-resolution using very deep convolutional networks},
  author={Kim, Jiwon and Lee, Jung Kwon and Lee, Kyoung Mu},
  booktitle={Proceedings of the IEEE conference on computer vision and pattern recognition},
  pages={1646--1654},
  year={2016}
}

@inproceedings{tai2017memnet,
  title={Memnet: A persistent memory network for image restoration},
  author={Tai, Ying and Yang, Jian and Liu, Xiaoming and Xu, Chunyan},
  booktitle={Proceedings of the IEEE International conference on computer vision},
  pages={4539--4547},
  year={2017}
}

@inproceedings{ledig2017photo,
  title={Photo-realistic single image super-resolution using a generative adversarial network},
  author={Ledig, Christian and Theis, Lucas and Husz{\'a}r, Ferenc and Caballero, Jose and Cunningham, Andrew and Acosta, Alejandro and Aitken, Andrew and Tejani, Alykhan and Totz, Johannes and Wang, Zehan and others},
  booktitle={Conference on computer vision and pattern recognition},
  pages={4681--4690},
  year={2017}
}

@inproceedings{yue2022blind,
  title={Blind image super-resolution with elaborate degradation modeling on noise and kernel},
  author={Yue, Zongsheng and Zhao, Qian and Xie, Jianwen and Zhang, Lei and Meng, Deyu and Wong, Kwan-Yee K},
  booktitle={Proceedings of the IEEE/CVF Conference on Computer Vision and Pattern Recognition},
  pages={2128--2138},
  year={2022}
}

@article{anwar2020deep,
  title={A deep journey into super-resolution: A survey},
  author={Anwar, Saeed and Khan, Salman and Barnes, Nick},
  journal={ACM Computing Surveys (CSUR)},
  volume={53},
  number={3},
  pages={1--34},
  year={2020},
  publisher={ACM New York, NY, USA}
}

@misc{kastryulin2022piq,
  title = {PyTorch Image Quality: Metrics for Image Quality Assessment},
  url = {https://arxiv.org/abs/2208.14818},
  author = {Kastryulin, Sergey and Zakirov, Jamil and Prokopenko, Denis and Dylov, Dmitry V.},
  doi = {10.48550/ARXIV.2208.14818},
  publisher = {arXiv},
  year = {2022}
}

@inproceedings{shi2016real,
  title={Real-time single image and video super-resolution using an efficient sub-pixel convolutional neural network},
  author={Shi, Wenzhe and Caballero, Jose and Husz{\'a}r, Ferenc and Totz, Johannes and Aitken, Andrew P and Bishop, Rob and Rueckert, Daniel and Wang, Zehan},
  booktitle={Proceedings of the IEEE conference on computer vision and pattern recognition},
  pages={1874--1883},
  year={2016}
}

\endgroup
}

\end{document}